\documentclass[aps,prl,twocolumn,superscriptaddress,showpacs]{revtex4}

\usepackage[dvips]{graphics}
\usepackage{bm,color}

\definecolor{red}{rgb}{1,0,0}
\definecolor{green}{rgb}{0,1,0}

\begin{document}

\title{Chirality distribution and transition energies of carbon nanotubes}

\author{H. Telg}
\affiliation{Institut f\"ur Festk\"orperphysik, Technische
Universit\"at Berlin, Hardenbergstr. 36, 10623 Berlin, Germany}

\author{J. Maultzsch}
\affiliation{Institut f\"ur Festk\"orperphysik, Technische
Universit\"at Berlin, Hardenbergstr. 36, 10623 Berlin, Germany}

\author{S. Reich}
\affiliation{Department of Engineering, University of Cambridge, Cambridge CB2 1PZ, United Kingdom}

\author{F. Hennrich}
\affiliation{Institut f\"ur Nanotechnologie, Forschungszentrum Karlsruhe, 76021 Karlsruhe, Germany}

\author{C. Thomsen}
\affiliation{Institut f\"ur Festk\"orperphysik, Technische
Universit\"at Berlin, Hardenbergstr. 36, 10623 Berlin, Germany}

\date{\today}

\begin{abstract}
From resonant Raman scattering on isolated nanotubes we obtained
the optical transition energies, the radial breathing mode
frequency and Raman intensity of both metallic and semiconducting
tubes. We unambiguously assigned the chiral index $(n_1,n_2)$ of
$\approx$ 50 nanotubes based solely on a third-neighbor
tight-binding Kataura plot and find
$\omega_{\mathrm{RBM}}=(214.4\pm 2)$\,cm$^{-1}$nm$/d+(18.7\pm
2)$\,cm$^{-1}$. In contrast to luminescence experiments we observe
all chiralities including zig-zag tubes. The Raman intensities
have a systematic chiral-angle dependence confirming recent
\emph{ab-initio} calculations.

\end{abstract}


\maketitle


The successful preparation of single-walled carbon nanotubes in
solution where the tubes are prevented from rebundling has opened
a new direction in carbon nanotube
research~\cite{connell02,bachilo02,lefebvre03,lebedkin03}. Strong luminescence by direct recombination from the band
gap was detected in these isolated tubes, whereas in
nanotube bundles no luminescence is observed. The
electronic structure of carbon nanotubes and  the optical
transition energies  vary  strongly with their chiral index
$(n_1,n_2)$~\cite{reich04}. Because the synthesis of nanotubes
with a predefined chiral index  has not  been achieved so far,
luminescence experiments were carried out on tube ensembles with
unknown composition of chiral angles. Several attempts to  assign
the chiral index $(n_1,n_2)$ to the experimentally observed
luminescence peaks were
reported~\cite{bachilo02,lebedkin03,hagen03,weisman03}. With a
unique assignment, one could validate and possibly revise
theoretical models of the electronic band structure. Moreover,
such an assignment would allow to characterize the tubes after their 
production and to control their separation~\cite{krupke03}.

Bachilo~\emph{et al.} suggested an $(n_1,n_2)$ assignment of the first and second
transition energies in semiconducting tubes~\cite{bachilo02}.
Their assignment is based on pattern recognition between
experiment and theory in a plot of the second transition (excitation energy)  \emph{versus} the first transition (emission energy)~\cite{reich02b}. The
patterns, however, were not unique, and the frequency of  the radial breathing mode (RBM) was  used to find an anchoring element
that singles out one of  the
assignments. Surprisingly, zig-zag tubes were not detected in these luminescence
experiments. Bachilo~\emph{et al.}  concluded that the concentration of tubes
with chiral angles close to the zig-zag direction was very low in
the sample~\cite{bachilo02}.

The electronic transition energies of metallic nanotubes cannot be
detected by luminescence experiments. An elegant approach is to
record Raman resonance profiles~\cite{jorio01a,canonico02,kramberger03,doorn04}, with maximum intensity close to the real
transitions in the electronic band structure. Resonance
profiles from nanotubes in solution were first reported by
Strano~\emph{et al.}~\cite{strano03b}; their  $(n_1,n_2)$ assignment to the
transition energies was based on the
RBM frequency to tube diameter relationship of Ref.~\cite{bachilo02}. The resonance
profiles of the so-assigned RBM peaks were then used to find an
empirical expression for the transition energies in metallic
tubes.

In this paper we present the transition energies of both metallic
and semiconducting nanotubes by resonant 
Raman spectroscopy.  Plotting the resonance maxima as a function
of inverse RBM frequency, we obtain an $(n_1,n_2)$ assignment
without any additional assumptions. From our assignment we fit 
$c_1=214.4$\,cm$^{-1}$nm and $c_2=18.7$\,cm$^{-1}$ for the
relation between diameter and RBM frequency. 
We observed several semiconducting tubes that were not detectable
by luminescence.
 Our results show that the
electron-phonon coupling strength increases systematically for
smaller chiral angles.
Conclusions about the distribution of chiral angles in a sample based solely on
luminescence intensity lead to incorrect results; in particular, zig-zag tubes are present in nanotube ensembles.

\begin{figure*}
\resizebox{1\textwidth}{!}{
\includegraphics*{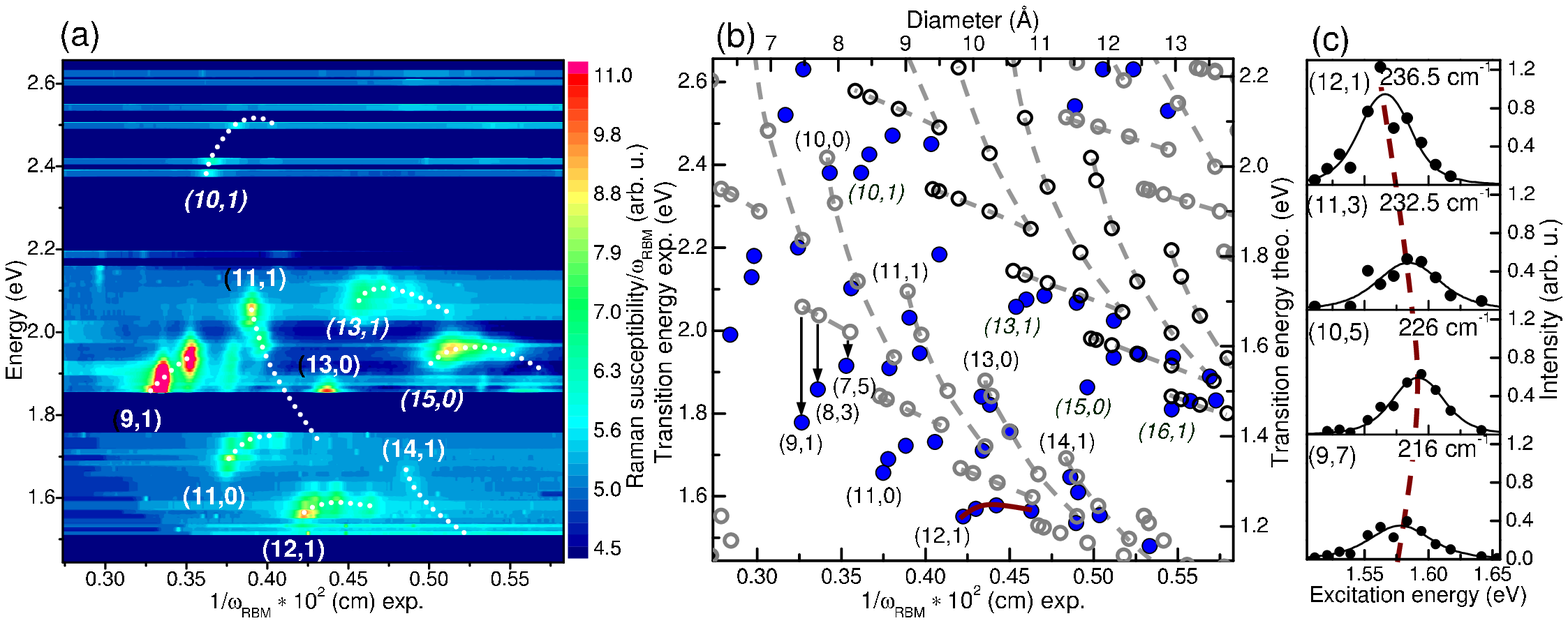}}
\caption{(Color online) (a) Contour plot of the Raman cross section of
the RBM as a function of excitation energy and reciprocal RBM
frequency. In (a) and (b) the dotted and dashed lines connect
maxima originating from tubes of the same branch. In each branch
the member with the largest RBM frequency is labeled. (b) Kataura plot
from experimental results (filled circles) and third-neighbor
tight-binding calculations (open circles). Gray and black indicate
semiconducting (2nd and 3rd transitions) and metallic tubes, respectively.
The plot consists of theoretical (right, top axes) and
experimental data (left, bottom axes). The error in experimental transition
energies is mostly smaller than 0.02\,eV. For the (9,1) branch,
the vertical arrows indicate for each member the assignment to its
theoretically predicted point.  The increasing softening compared
to the theoretical transition energies with smaller chiral angle is clearly seen. (c) Integrated Raman intensity as a
function of excitation energy for the tubes in the (12,1) branch. } \label{FehlKataurRes}
\end{figure*}

We performed Raman spectroscopy on HiPCO nanotubes with diameters $d\approx0.7 - 1.2$\,nm~\cite{nikolaev99}. We dispersed the 
tubes in D$_2$O containing a surfactant, see 
Ref.~\cite{lebedkin03} and used  an Ar-Kr laser
between 2.18 and 2.62\,eV and two tunable lasers (1.85-2.15\,eV and 1.51-1.75\,eV). The  spectra were collected with a Dilor XY800 spectrometer in
backscattering geometry at room temperature. 
To obtain the Raman cross section from the measured intensity we normalized the spectra to CaF$_2$ and BaF$_2$ measurements taken under the same experimental conditions (integration time, laser power). This also corrects for the spectrometer sensitivity and the $\omega^4$ dependence of the Raman process. The Raman susceptibility was calculated from the normalized spectra by dividing by the Bose-Einstein occupation number and the inverse phonon frequency~\cite{cardona82}. The latter was omitted in Fig.~\ref{FehlKataurRes}\,(a) for a better
representation.

In Fig.~\ref{FehlKataurRes}\,(a) we show a contour plot of all
Raman spectra, \emph{i.e.}, the Raman scattering power as a
function of inverse RBM frequency ($1/\omega_{\mathrm{RBM}}$) and  excitation energy.
When tuning the excitation energy, the RBM peaks
appear and disappear in groups of close-by frequencies. These are
indicated by the dotted lines, where the later-assigned chiral
index of the largest RBM frequency of each group is given. To
obtain the optical transition energies, we use the resonance profiles, see, \emph{e.g.}, Fig.~\ref{FehlKataurRes}\,(c) for the group beginning with the (12,1) tube. The
lines are fits to the first-order Raman cross section including incoming and outgoing  resonance. From these fits we  obtain directly the transition energies, which are by $\approx10$\,meV ($0.5\,\omega_{\mathrm{RBM}}$) smaller than the energy of the resonance maximum. For
decreasing RBM frequency, the resonance energy first increases and
then decreases again [dashed line in
Fig.~\ref{FehlKataurRes}\,(c)].  The maximum intensity
decreases with decreasing RBM frequency. The described systematics
are valid only for RBMs within a given group of tubes  defined
by the dotted lines in Fig.~\ref{FehlKataurRes}\,(a).

In Fig.~\ref{FehlKataurRes}\,(b) we plot the experimental transition energies as a function of $1/\omega_{\mathrm{RBM}}$ (solid
circles, bottom and left axes). Since  $1/\omega_{\mathrm{RBM}}$
is approximately proportional to the tube diameter, we thereby
obtain an experimental Kataura plot, which  we compare  to
third-nearest neighbor tight-binding calculations~\cite{reich02b}
(open circles, upper and right  axes). For semiconducting tubes,
the measured transition energies are in excellent agreement with
the $E_{22}$ energies measured by Bachilo \emph{et
al.}~\cite{bachilo02}. From a comparison between the experimental
and the calculated Kataura plot we were able to assign the tube
chiralities in the following way. Since we did not want to make
any assumptions about the coefficients $c_1$ and $c_2$ in
the diameter-RBM frequency relationship,
$\omega_{\mathrm{RBM}}=(c_1/d+c_2)$, we varied  $c_1$
and $c_2$ to find the best match between experimental and
calculated data in Fig.~\ref{FehlKataurRes}\,(b). Varying $c_1$
and $c_2$, respectively, corresponds to stretching and displacing
the theoretical Kataura plot along the $1/\omega_{\mathrm{RBM}}$  axis. In addition
we displaced the plot on the energy scale. The origin of this
overall energy offset is, among others, found in the tight-binding
parameters, which were fitted to LDA calculations believed to
underestimate the transition energies by typically 10~\%.

 Figure~\ref{FehlKataurRes}\,(b) shows the best $(n_1,n_2)$ assignment
of  experimental transition energies and RBM frequencies. The
dashed lines in the theoretical Kataura plot indicate branches
that are formed by tubes with $2n_1+n_2$ constant.  The indices of
neighboring tubes in a given branch  are related  through $(n_1+1,n_2-2)$ or $(n_1-1,n_2+2)$, as long as $n_1\geq n_2$. For example, the branch
starting with the (11,0) tube contains  from small to large
diameter (11,0), (10,2), (9,4), and (8,6), and belongs to the
$\nu=(n_1-n_2)\mathrm{mod\,}3=-1$ family. The chiral angle $\theta$ of
the tubes within a branch increases towards larger diameter.
Zig-zag tubes ($\theta =0^{\circ}$) are  always at the outermost
position of a branch. The  RBM groups in
Fig.~\ref{FehlKataurRes}\,(a) [dotted lines]  directly correspond to 
these branches, see, \emph{e.g.}, the one beginning with the
(9,1) tube.

We find a very good match between the patterns of the calculated diameter and of the experimental
values  $1/\omega_{\mathrm{RBM}}$. The experimental
transition energies, however, deviate systematically from the calculations
 for  the branches on the low-energy side of the
semiconducting and of the metallic transitions. 
This deviation increases for smaller chiral angles.
In other words, the experimental transition energies bend
downwards from the calculated transitions with decreasing chiral
angle. Such a strong deviation between theory and experiment is
not seen for the upper Kataura branches, where the tubes belong to
the $\nu =+1$ family. This was  observed in
luminescence experiments as well and will be discussed below.

We considered alternative assignments by
displacing the experimental data along the sets of semiconducting
and metallic tubes in Fig.~\ref{FehlKataurRes}\,(b) to the left and to
the right. None of them yields a good agreement regarding  the
data points within the $(n_1-1,n_2+2)$ branches: Either the \emph{number}
of tubes differs between the theoretical and experimental branch
or some points are eminently displaced \emph{horizontally} from the
theoretical ones. Having found the zig-zag tubes we can exclude these alternative assignments.
Table~\ref{assign} summarizes the assignment of chiral indices to the measured RBM frequencies and transition energies for the branch of the (11,0) tube~\cite{langetabelle}.

\begin{table}
\caption{\label{assign}Measured $\omega_{\mathrm{RBM}}$ and $E_{22}$ for the branch of the (11,0) tube. See also supplementary material~\cite{langetabelle}.}
\begin{ruledtabular}
\begin{tabular}{lccccccc}
 chiral index&(11,0) &(10,2) &(9,4) & (8,6) \\ \hline
$\omega_{\mathrm{RBM}}$ (cm$^{-1}$) & 266.7 &264.6 &257.5 &246.4 \\
$E_{22}$ (eV) & 1.657& 1.690& 1.72 & 1.73 
\end{tabular}
\end{ruledtabular}
\end{table}


From Fig.~\ref{FehlKataurRes}\,(b) we can now fit the
coefficients of the linear relationship between  $\omega_\mathrm{RBM}$ and $1/d$
 of the assigned nanotubes. The linear fit
based on 45 identified tubes yields
$\omega_\mathrm{RBM}=c_1/d+c_2$ with $c_1=(214.4\pm
2)$\,cm$^{-1}$nm and $c_2=(18.7\pm 2)$\,cm$^{-1}$. 
Here, the tube diameter is  geometrically
determined by $d=a_0\sqrt{n_1^2+n_1n_2+n_2^2}/\pi$, using a
graphite lattice constant $a_0=2.461$\,\AA. 
 $c_1$ and $c_2$ differ
somewhat from the coefficients found in Ref.~\cite{bachilo02} for
the same type of samples, because we used many more RBM
frequencies for the fit and calculated the diameter from a smaller
$a_0$.
The coefficients are similar to theoretical predictions
\cite{kuerti98,dobardzic03} but different from other experimental
work~\cite{jorio01,kramberger03}. 
In Ref.~\cite{kramberger03},  $c_1$ and $c_2$ were used as free parameters to \emph{find} the assignment; moreover,  the experimental information on the transition energies was not included. In Ref.~\cite{jorio01}, the chiral index assignment also depends on $c_1$ ($c_2=0$). Moreover, the tubes are on a substrate, which might alter the RBM frequencies. 
In contrast, our assignment is not based on a choice for   $c_1$ and $c_2$. For the first time, they are  obtained by a linear fit \emph{after} the assignment was performed. For example, the (11,0) tube has an RBM frequency of 266.7\,cm$^{-1}$ and $E_{22}=1.657$\,eV regardless of its exact diameter or any fitting procedure, see Table~\ref{assign}.

In addition to the  semiconducting tubes, we directly obtained
the transition energies of metallic tubes.
Our assignment of the metallic tubes to RBM frequencies agrees well with Strano \emph{et al.}~\cite{strano03b}.
The empirically obtained transition energies in
Ref.~\cite{strano03b}, however,   underestimate the
experimental  values in Fig.~\ref{FehlKataurRes}\,(b). Moreover, our
data show a  stronger bending of the branches towards small
chiral angles. This discrepancy  comes mainly from the presence of
pairs of close-by transition energies in chiral metallic
tubes~\cite{reich00c}. This ambiguity led to an incorrect
assignment of single data points to the upper or lower transition.

Given the unique ($n_1,n_2$) assignment of the RBM and the resonance energies, we can now examine the chirality dependence
of the transition energies and of the RBM intensity. The optical transition energies of carbon nanotubes are
roughly proportional to $1/d$~\cite{mintmire98,kataura99,reich00c,reich02b}. Tight-binding
calculations predict deviations from a
pure $1/d$ dependence as a function of chiral
angle~\cite{reich00c}. They lead to the
branches in the Kataura plot and are more pronounced if third-nearest
neighbors are included~\cite{reich02b}. Still,
the band gap energies of tubes with small chiral angle (zig-zag
tubes) are systematically lower in first-principles calculations
than in zone folding~\cite{reich02}. The
branches in the experimental Kataura plot of
Fig.~\ref{FehlKataurRes}\,(b) are an experimental verification of the
predictions from \emph{ab-initio} calculations.

The chiral-angle dependent softening of the transition energies
with respect to the third-order tight-binding calculations is due
to rehybridization of the $\pi$ and $\sigma$
bands~\cite{blase94,reich02,reich02b}. It is stronger for the
states originating from between the $K$ and $M$ point of the
graphite Brillouin zone in the zone-folding approach and weaker
for states from the other side of the $K$ point~\cite{reich02}. In semiconducting tubes, the
value of $\nu=(n_1-n_2)\,\textrm{mod }3$ determines from which side of
the $K$ point the electronic states originate for a given optical
transition. The tubes in the lower branches of the $E_{22}$ 
transitions in semiconducting tubes, \emph{i.e.}, of the branches
beginning with (9,1), (11,0), and (12,1) in
Fig.~\ref{FehlKataurRes}\,(b),  have $\nu=-1$. The softening for small-chiral-angle tubes in these
branches compared to the tight-binding value is quite strong. The
tubes in the upper part of the same set of transitions have
$\nu=+1$  and are less affected  by
the softening. For example, the experimental transition energy of
the (10,0) tube [$\theta=0^{\circ}$, $\nu=+1$] matches the theoretical value very well 
[Fig.~\ref{FehlKataurRes}\,(b)]. Taking this softening of the
transition energies in the $-1$ families into account, the
agreement of our data with the theoretical predictions is
excellent.

In general, the RBM signal was strong for nanotubes with
$\nu=-1$ and also from the lower branches of
the metallic transitions in Fig.~\ref{FehlKataurRes}\,(b). The
intensities were by a factor of four to ten weaker for tubes with
$\nu=+1$. This observation confirms
\emph{ab-initio} calculations of the electron-phonon coupling that
predicted the magnitude of the electron-phonon matrix element to
alternate with  $\nu$~\cite{machon03}.  The relative Raman intensity of the RBM can also be used to discriminate between the two families of semiconducting tubes.


\begin{figure}
\resizebox{1\columnwidth}{!}{
\includegraphics*{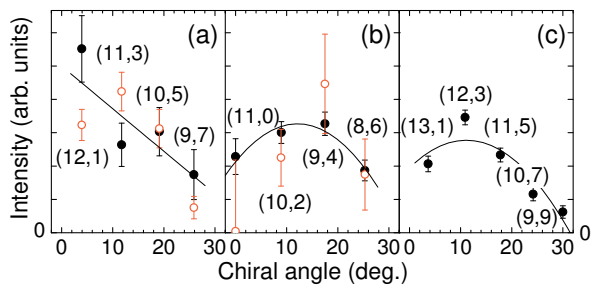}}
\caption{(Color online) Raman intensity as a function of the
chiral angle for three nanotube branches (solid circles) with
$(n_1,n_2)$ as indicated. Open circles: calculated Raman
intensity, see text. (a) and (b) contain semiconducting, (c)
contains metallic tubes. \label{chiralIntensity}}
\end{figure}

The assignment of the semiconducting tubes in
Fig.~\ref{FehlKataurRes}\,(b) corresponds to the one found by
Bachilo \emph{et al.} from luminescence
experiments~\cite{bachilo02}. In contrast to the luminescence
results, which reported a maximum intensity for close-to-armchair
tubes and no emission from zig-zag tubes, we clearly observed
zig-zag or close-to-zig-zag tubes as well. The (13,0), (11,0), and
the (10,0) tube show that zig-zag tubes are present in the sample.
These  tubes as well as the (14,1) tube were not observed by 
photoluminescence. An important conclusion from our work is
that the absence of photoluminescence from these tubes thus does
not imply the preferential growth of armchair tubes.

In Fig.~\ref{chiralIntensity} we show the resonance maxima for tubes belonging to the
same $(n_1-1,n_2+2)$ branch. The Raman intensity increases with decreasing
chiral angle $\theta$ and is at maximum for $\theta\approx10-15^\circ$, except for
(a). Theoretically, the Raman  amplitude is proportional to the
electron-phonon coupling times the square of the optical absorption
strength~\cite{cardona82}. Mach\'on~\emph{et al.}~\cite{machon03} found that
the electron-phonon coupling of the RBM decreases strongly with increasing
$\theta$. We therefore model the electron-phonon interaction by a linear
function of the chiral angle with a three times stronger coupling for
zig-zag than for armchair tubes as found in Ref.~\cite{machon03}. We approximate the optical absorption strength of a tube
 by its experimental photoluminescence
intensity~\cite{bachilo02}, \emph{i.e.}, we assume the absorption and

emission probability to be the same. The relative Raman intensities
calculated with this model are in good agreement with experiment (open dots in
Fig.~\ref{chiralIntensity}). The Raman response of zig-zag tubes is enhanced compared
to their luminescence intensity because of their strong electron-phonon
coupling. On the other hand, the Raman intensity of zig-zag tubes is
smaller than for $\theta\approx10^\circ$ due to the small absorption
coefficient.
Thus, our data are completely
consistent with a uniform chirality distribution in the sample.

In conclusion, we assigned the chiral indices to $\approx 50$ measured RBM frequencies and transition energies by resonant Raman spectroscopy. In contrast to all previous  work our assignment is independent of  the coefficients $c_1$ and $c_2$, which we fit only \emph{after} assigning the chiral index to a particular RBM. 
The largest Raman
intensity was measured for tubes with chiral angles around $15^{\circ}$
or smaller, which is in agreement with theoretical predictions
and implies that the chiralities are evenly distributed. Moreover,
our results confirm that the RBM intensity in semiconducting tubes
depends on the $(n_1-n_2)\,\textrm{mod }3$ family. The transition
energies deviate from  zone-folding predictions with decreasing
chiral angle, which, in particular for metallic tubes, was
strongly underestimated in earlier work.

\begin{acknowledgments}
This work was supported  by the DFG
under grant number Th662/8-2. S. R. was supported by the Oppenheimer Fund
and  Newnham College.
\end{acknowledgments}


\begin{thebibliography}{10}
\expandafter\ifx\csname bibnamefont\endcsname\relax
  \def\bibnamefont#1{#1}\fi
\expandafter\ifx\csname bibfnamefont\endcsname\relax
  \def\bibfnamefont#1{#1}\fi
\expandafter\ifx\csname url\endcsname\relax
  \def\url#1{\texttt{#1}}\fi
\expandafter\ifx\csname urlprefix\endcsname\relax\def\urlprefix{URL }\fi
\expandafter\ifx\csname bibinfo\endcsname\relax \def\bibinfo#1#2{#2}\fi
\expandafter\ifx\csname eprint\endcsname\relax \def\eprint#1{#1}\fi

\bibitem{connell02}
\bibinfo{author}{\bibfnamefont{M.~J.} \bibnamefont{O{'}Connell}},
  \emph{et~al.},
  \bibinfo{journal}{Science} \textbf{\bibinfo{volume}{297}},
  \bibinfo{pages}{593} (\bibinfo{year}{2002}).

\bibitem{bachilo02}
\bibinfo{author}{\bibfnamefont{S.~M.} \bibnamefont{Bachilo}},
\emph{et~al.},  
\bibinfo{journal}{Science}
  \textbf{\bibinfo{volume}{298}}, \bibinfo{pages}{2361} (\bibinfo{year}{2002}).

\bibitem{lefebvre03}
\bibinfo{author}{\bibfnamefont{J.}~\bibnamefont{Lefebvre}},
  \emph{et~al.},
  \bibinfo{journal}{Phys. Rev. Lett.} \textbf{\bibinfo{volume}{90}},
  \bibinfo{pages}{217401} (\bibinfo{year}{2003}).

\bibitem{lebedkin03}
\bibinfo{author}{\bibfnamefont{S.}~\bibnamefont{Lebedkin}},
  \emph{et~al.},
  \bibinfo{journal}{J. Phys. Chem. B} \textbf{\bibinfo{volume}{107}},
  \bibinfo{pages}{1949} (\bibinfo{year}{2003}).

\bibitem{reich04}
\bibinfo{author}{\bibfnamefont{S.}~\bibnamefont{Reich}},
  \bibinfo{author}{\bibfnamefont{C.}~\bibnamefont{Thomsen}}, \bibnamefont{and}
  \bibinfo{author}{\bibfnamefont{J.}~\bibnamefont{Maultzsch}},
  \emph{\bibinfo{title}{Carbon {N}anotubes: {B}asic {C}oncepts and {P}hysical
  {P}roperties}} (\bibinfo{publisher}{Wiley-VCH}, \bibinfo{address}{Berlin},
  \bibinfo{year}{2004}).

\bibitem{hagen03}
\bibinfo{author}{\bibfnamefont{A.}~\bibnamefont{Hagen}} \bibnamefont{and}
  \bibinfo{author}{\bibfnamefont{T.}~\bibnamefont{Hertel}},
  \bibinfo{journal}{Nano Lett.} \textbf{\bibinfo{volume}{3}},
  \bibinfo{pages}{383} (\bibinfo{year}{2003}).

\bibitem{weisman03}
\bibinfo{author}{\bibfnamefont{R.~B.} \bibnamefont{Weisman}} \bibnamefont{and}
  \bibinfo{author}{\bibfnamefont{S.~M.} \bibnamefont{Bachilo}},
  \bibinfo{journal}{Nano Lett.} \textbf{\bibinfo{volume}{3}},
  \bibinfo{pages}{1235} (\bibinfo{year}{2003}).

\bibitem{krupke03}
\bibinfo{author}{\bibfnamefont{R.}~\bibnamefont{Krupke}},
  \bibinfo{author}{\bibfnamefont{F.}~\bibnamefont{Hennrich}},
  \bibinfo{author}{\bibfnamefont{H.}~\bibnamefont{v.~L{\"o}hneysen}},
  \bibnamefont{and} \bibinfo{author}{\bibfnamefont{M.~M.}
  \bibnamefont{Kappes}}, \bibinfo{journal}{Science}
  \textbf{\bibinfo{volume}{301}}, \bibinfo{pages}{344} (\bibinfo{year}{2003}).

\bibitem{reich02b}
\bibinfo{author}{\bibfnamefont{S.}~\bibnamefont{Reich}},
  \bibinfo{author}{\bibfnamefont{J.}~\bibnamefont{Maultzsch}},
  \bibinfo{author}{\bibfnamefont{C.}~\bibnamefont{Thomsen}}, \bibnamefont{and}
  \bibinfo{author}{\bibfnamefont{P.}~\bibnamefont{Ordej{\' o}n}},
  \bibinfo{journal}{Phys. Rev. B} \textbf{\bibinfo{volume}{66}},
  \bibinfo{pages}{035412} (\bibinfo{year}{2002}).

\bibitem{jorio01a}
\bibinfo{author}{\bibfnamefont{A.}~\bibnamefont{Jorio}},
 \emph{et~al.},
\bibinfo{journal}{Phys. Rev. B}
  \textbf{\bibinfo{volume}{63}}, \bibinfo{pages}{245416}
  (\bibinfo{year}{2001}).

\bibitem{canonico02}
\bibinfo{author}{\bibfnamefont{M.}~\bibnamefont{Canonico}},
  \emph{et~al.},
  \bibinfo{journal}{Phys. Rev. B} \textbf{\bibinfo{volume}{65}},
  \bibinfo{pages}{201402(R)} (\bibinfo{year}{2002}).

\bibitem{kramberger03}
\bibinfo{author}{\bibfnamefont{C.}~\bibnamefont{Kramberger}},
  \emph{et~al.},
  \bibinfo{journal}{Phys. Rev. B} \textbf{\bibinfo{volume}{68}},
  \bibinfo{pages}{235404} (\bibinfo{year}{2003}).

\bibitem{doorn04}
\bibinfo{author}{\bibfnamefont{S.~K.} \bibnamefont{Doorn}},
  \emph{et~al.},
  \bibinfo{journal}{Appl. Phys. A} \textbf{\bibinfo{volume}{78}},
  \bibinfo{pages}{1147} (\bibinfo{year}{2004}).

\bibitem{strano03b}
\bibinfo{author}{\bibfnamefont{M.~S.} \bibnamefont{Strano}},
  \emph{et~al.},
  \bibinfo{journal}{Nano Lett.} \textbf{\bibinfo{volume}{3}},
  \bibinfo{pages}{1091} (\bibinfo{year}{2003}).

\bibitem{nikolaev99}
\bibinfo{author}{\bibfnamefont{P.}~\bibnamefont{Nikolaev}},
  \emph{et~al.},
  \bibinfo{journal}{Chem. Phys. Lett.} \textbf{\bibinfo{volume}{313}},
  \bibinfo{pages}{91} (\bibinfo{year}{1999}).

\bibitem{cardona82}
\bibinfo{author}{\bibfnamefont{M.}~\bibnamefont{Cardona}}, in
  \emph{\bibinfo{booktitle}{Light Scattering in Solids II}}, edited by
  \bibinfo{editor}{\bibfnamefont{M.}~\bibnamefont{Cardona}} \bibnamefont{and}
  \bibinfo{editor}{\bibfnamefont{G.}~\bibnamefont{G{\"u}ntherodt}}
  (\bibinfo{publisher}{Springer}, \bibinfo{address}{Berlin},
  \bibinfo{year}{1982}), vol.~\bibinfo{volume}{50} of
  \emph{\bibinfo{series}{Topics in Applied Physics}}, p.~\bibinfo{pages}{19}.

\bibitem{langetabelle}
 \bibinfo{note}{See EPAPS Document No. [] for a table containing the complete
  branches of Fig.~\ref{FehlKataurRes}\,(b).}

\bibitem{kuerti98}
\bibinfo{author}{\bibfnamefont{J.}~\bibnamefont{K{\"u}rti}},
  \bibinfo{author}{\bibfnamefont{G.}~\bibnamefont{Kresse}}, \bibnamefont{and}
  \bibinfo{author}{\bibfnamefont{H.}~\bibnamefont{Kuzmany}},
  \bibinfo{journal}{Phys. Rev. B} \textbf{\bibinfo{volume}{58}},
  \bibinfo{pages}{8869} (\bibinfo{year}{1998}).

\bibitem{dobardzic03}
\bibinfo{author}{\bibfnamefont{E.}~\bibnamefont{Dobard{\v z}i{\' c}}},
  \bibinfo{author}{\bibfnamefont{I.~M.~B.} \bibnamefont{Nikoli{\'c}}},
  \bibinfo{author}{\bibfnamefont{T.}~\bibnamefont{Vukovi{\'c}}},
  \bibnamefont{and}
  \bibinfo{author}{\bibfnamefont{M.}~\bibnamefont{Damnjanovi{\'c}}},
  \bibinfo{journal}{Phys. Rev. B} \textbf{\bibinfo{volume}{68}},
  \bibinfo{pages}{045\,408} (\bibinfo{year}{2003}).

\bibitem{jorio01}
\bibinfo{author}{\bibfnamefont{A.}~\bibnamefont{Jorio}},
 \emph{et~al.},
\bibinfo{journal}{Phys. Rev. Lett.}
  \textbf{\bibinfo{volume}{86}}, \bibinfo{pages}{1118} (\bibinfo{year}{2001}).

\bibitem{reich00c}
\bibinfo{author}{\bibfnamefont{S.}~\bibnamefont{Reich}} \bibnamefont{and}
  \bibinfo{author}{\bibfnamefont{C.}~\bibnamefont{Thomsen}},
  \bibinfo{journal}{Phys. Rev. B} \textbf{\bibinfo{volume}{62}},
  \bibinfo{pages}{4273} (\bibinfo{year}{2000}).

\bibitem{mintmire98}
\bibinfo{author}{\bibfnamefont{J.~W.} \bibnamefont{Mintmire}} \bibnamefont{and}
  \bibinfo{author}{\bibfnamefont{C.~T.} \bibnamefont{White}},
  \bibinfo{journal}{Phys. Rev. Lett.} \textbf{\bibinfo{volume}{81}},
  \bibinfo{pages}{2506} (\bibinfo{year}{1998}).

\bibitem{kataura99}
\bibinfo{author}{\bibfnamefont{H.}~\bibnamefont{Kataura}},
  \emph{et~al.},
  \bibinfo{journal}{Synth. Met.} \textbf{\bibinfo{volume}{103}},
  \bibinfo{pages}{2555} (\bibinfo{year}{1999}).

\bibitem{reich02}
\bibinfo{author}{\bibfnamefont{S.}~\bibnamefont{Reich}},
  \bibinfo{author}{\bibfnamefont{C.}~\bibnamefont{Thomsen}}, \bibnamefont{and}
  \bibinfo{author}{\bibfnamefont{P.}~\bibnamefont{Ordej{\' o}n}},
  \bibinfo{journal}{Phys. Rev. B} \textbf{\bibinfo{volume}{65}},
  \bibinfo{pages}{155411} (\bibinfo{year}{2002}).

\bibitem{blase94}
\bibinfo{author}{\bibfnamefont{X.}~\bibnamefont{Blase}},
 \emph{et~al.},
  \bibinfo{journal}{Phys. Rev. Lett.} \textbf{\bibinfo{volume}{72}},
  \bibinfo{pages}{1878} (\bibinfo{year}{1994}).

\bibitem{machon03}
\bibinfo{author}{\bibfnamefont{M.}~\bibnamefont{Mach{\'o}n}},
   \emph{et~al.},
  (\bibinfo{year}{2003}), \bibinfo{note}{submitted to Phys. Rev. B}; cond-mat/0408436v1.

\end{thebibliography}
\end{document}